\documentclass[structabstract]{aa}  
\usepackage{graphicx}
\usepackage{natbib}
\usepackage{txfonts}
\begin{document}
\title{Non-LTE models for the gaseous metal component of circumstellar discs around white dwarfs}

\author{S. Hartmann \and T. Nagel  \and  T. Rauch \and K. Werner}

\institute{Institute for Astronomy and Astrophysics, Kepler Center for Astro
  and Particle Physics, Eberhard Karls University T\"ubingen,
  Sand~1, 72076 T\"ubingen, Germany\\
  \email{hartmann@astro.uni-tuebingen.de}}

\date{Received xx.xx.xxxx; accepted xx.xx.xxxx}

\authorrunning{Hartmann et al.}
\titlerunning{Non-LTE models for gaseous circumstellar discs around white dwarfs}

\abstract
% context heading (optional), leave it empty if necessary
{Gaseous metal discs around single white dwarfs have been discovered recently. They are thought to develop from disrupted planetary bodies.}
{Spectroscopic analyses will allow us to study the composition of extrasolar planetary material. We investigate in detail the first object for which a gas disc was discovered (SDSS\,J122859.93+104032.9).}
{We perform non-LTE modelling of viscous gas discs by computing the detailed vertical structure and line spectra. The models are composed of carbon, oxygen, magnesium, silicon, calcium, and hydrogen with chemical abundances typical for Solar System asteroids. Line asymmetries are modelled by assuming spiral-arm and eccentric disc structures as suggested by hydrodynamical simulations.}
{The observed infrared Ca\,{\sc ii} emission triplet can be modelled with a hydrogen-deficient metal gas disc located inside of the tidal disruption radius, with $T_{\rm eff}$\,$\approx$\,6\,000\,K and a surface mass density of $\mathit{\Sigma}\approx$\,0.3\,g/cm$^2$.  The inner radius is well constrained at about 0.64\,R$_\odot$. The line profile asymmetry can be reproduced by either a spiral-arm structure or an eccentric disc, the latter being favoured by its time variability behaviour. Such structures, reaching from 0.64 to 1.5\,R$_\odot$, contain a mass of about 3--6\,$\cdot 10^{21}$\,g, the latter equivalent to the mass of a 135-km diameter Solar System asteroid.}
{}
    
\maketitle
%
%________________________________________________________________
\section{Introduction}

A significant fraction of white dwarfs (WD) that have cooled below $T_{\rm{eff}}\approx 25\,000$\,K (20--30\%) displays photospheric absorption lines from metals \citep{zuc10}. These polluted WDs must actively accrete matter at rates from $10^{-15}$ to $10^{-17}\,$M$_\odot$/yr, otherwise the atmospheres would have been purified by gravitational settling of heavy elements \citep{paq86,Koester:2006p1769,koe09}.  Until the recent past it was assumed that these stars accrete matter from the interstellar medium \citep{dup92}. This view has changed considerably in the past few years because it was discovered that many of the polluted WDs host dust discs within the stellar tidal radius \citep{Zuckerman:1987p710,Becklin:2005p721,Kilic:2005p907,Kilic:2006p908,Farihi:2007p715,Farihi:2009p718,Farihi:2010ApJ...714.1386F}. It is now widely accepted that WDs are polluted by matter accretion from these discs. It is thought that they contain material from tidally disrupted asteroids that were scattered towards the central stars as a consequence of dynamical resettling of a planetary system in the post-main sequence phase \citep{Debes:2002p712,Jura:2003p816}.

The photospheric metal abundance pattern in the polluted WDs allows us to \emph{indirectly} measure the composition of the accreted matter. This opens up the exciting possibility of studying the composition of extrasolar planetary material. The first impressive results have already been obtained, showing that the orbiting debris broadly mimics the terrestrial matter of the inner Solar System, including the possibility of water \citep{zuc07,Klein:2010p719,duf10,far11}. These results are based on our knowledge of metal diffusion rates in WD atmospheres and envelopes, which are difficult to obtain \citep{koe09}. Several uncertainties can affect the resulting composition of accreted material as concluded from the photospheric abundance pattern. One example is the assumption of stationary accretion on several diffusion timescales of all elements involved. It is therefore highly desirable to exploit alternative possibilities of determining the chemical composition of the accreted material.

Such an alternative method is offered by the recent discovery that three circumstellar dust discs around polluted WDs also host gaseous metal components that are interpreted as the collisional remains of solid material \citep{Gansicke:2006p444,Gansicke:2007p433,Gansicke:2008p430}. We are developing accretion disc models to derive the chemical abundances in the gas discs, so
that we should be able to \emph{directly} measure the composition of the parent planetary material. This method is complementary to the measurement of photospheric abundances because the disc spectra might reveal trace elements that are not seen in the WD spectra. It also provides a means to test our understanding of diffusion processes in the WD atmospheres and envelopes.  Building on our preliminary work \citep{Werner:2009p373}, we present here new results for our modelling efforts of gas discs.

In Sect.\,\ref{acdc}, we briefly introduce our method, followed by a description of the object that we study in detail (Sect.\,\ref{sdss}). In Sect.\,\ref{results}, we first summarise our results concerning simple models for pure-calcium discs, investigating the influence of effective temperature and surface-mass density on the emergent spectrum. We then present vertical structures and spectra of discs composed of an asteroid-like mixture of light metals, comprising C, O, Mg, Si, and Ca. Finally, we present results for non-axisymmetric disc geometries in order to explain the observed asymmetry of the double-peaked line profiles. Based on hydrodynamical simulations, we investigate the time evolution of this asymmetry in comparison with observations.

%__________________________________________________________________
\section{Accretion-disc modelling}\label{acdc}

For calculating geometrically thin accretion-disc models, we use our code \emph{AcDc} \citep{Nagel:2004p5}.  We assume axial symmetry, so that we can separate the disc into concentric rings of plane-parallel geometry. In that way, the radiative transfer becomes a one-dimensional problem. By integrating the spectra of the individual rings, we obtain a complete disc spectrum for different inclination angles.

The free parameters of one ring with radius $R$ are effective temperature $T_{\rm{eff}}(R)$, surface mass density $\mathit{\Sigma}(R)$, chemical composition, and the WD mass $M_{\rm{WD}}$. For the energy equation we assume that the emitted radiation is viscously generated, so the Reynolds number $Re$ (or $\alpha$) enters as an additional parameter. In the case of viscous $\alpha$-discs, the radial run of $T_{\rm{eff}}(R)$ and $\mathit{\Sigma}(R)$ can be expressed in terms of the mass-accretion rate and mass $M_{\rm{WD}}$ and radius $R_{\rm{WD}}$ of the central star (Shakura \& Sunyaev 1973). For comparison with observations, the emergent spectra from ring segments are Doppler-shifted to account for Keplerian motion, hence $R_{\rm{WD}}$ and disc inclination $i$ appear as additional parameters.

For each disc ring, the following set of coupled equations were solved simultaneously under the constraints of particle number and charge conservation:

\begin{itemize}

\item radiation transfer for the specific intensity $I$ at frequency $\nu$
\begin{equation}
\mu\,\frac{\partial\,I(\nu,\mu,z)}{\partial\,z}\,=\,-\chi(\nu,z)\,I(\nu,\mu,z)\,+\,\eta(\nu,z)
\end{equation}
with the absorption coefficient $\chi$, the emission coefficient $\eta$, the disc height $z$ above the midplane, and $\mu=\cos{\theta}$, with $\theta$ the angle between the ray and $z$;

\item hydrostatic equilibrium of gravitation, gas pressure $P_{\rm{gas}}$, and radiation pressure
\begin{equation}
\frac{\mathrm{d}P_{\rm{gas}}}{\mathrm{d}m}\,=\,\frac{G\,M_{\rm{WD}}}{R^3}\,z\,-\frac{4\pi}{c}\int\limits_{0}^{\infty}\frac{\chi(\nu)}{\rho}H(\nu,z)\,\mathrm{d}\nu
\end{equation}
with $\rho$ denoting the mass density and $H$ the Eddington flux. Here, we also introduced the column-mass density $m$ as   
\begin{equation}
m(z)\,=\,\int\limits_{z}^{\infty}\rho(z^\prime)\,\mathrm{d}z^\prime;
\end{equation}

\item energy balance between the viscously generated energy $E_{\rm{mech}}$ and the radiative energy loss $E_{\rm{rad}}$
\begin{equation}
E_{\rm{mech}}\,=\,E_{\rm{rad}}
\end{equation}
with
\begin{equation}\label{eqmech}
E_{\rm{mech}}\,=\,w\,\mathit{\Sigma}\,\left(R\,\frac{\mathrm{d}\omega}{\mathrm{d}R}\right)^2\,=\,\frac{9}{4}\,w\,\mathit{\Sigma}\,\frac{G\,M_{\rm{WD}}}{R^3}
\end{equation}
and
\begin{equation}
E_{\rm{rad}}\,=\,4\,\pi\,\int\limits_{0}^{\infty}\left[\eta(\nu,z)\,-\,\chi(\nu,z)\,J(\nu,z)\,\right]\,\mathrm{d}\nu
\end{equation}
with the angular velocity $\omega$, the mean intensity $J$, and $w$ the kinematic viscosity written following \cite{LB74}:
\begin{equation}
w=\sqrt{G\,M_{\rm{WD}}\,R}\,/\,Re.
\end{equation}
 For the models presented here we assume  $Re\,=\,15\,000$;

\item NLTE rate equations for the population numbers $n_i$ of the atomic levels $i$
\begin{equation}
n_i\sum_{i\neq j}P_{ij}\,-\,\sum_{j\neq i}n_j\,P_{ji}\,=\,0,
\end{equation}
where $P_{ij}$ denotes the rate coefficients, consisting of radiative and electron collisional components.

\end{itemize}

\begin{table}
\caption{Statistics of the model atoms used in our disc models.}   
\label{tab1}
\centering      
\begin{tabular}{l r r r} 
\hline\hline                
Ion & LTE levels & NLTE levels & Lines \\ 
\hline
H\,{\sc i}      &    6   &    10   &    45   \\ 
H\,{\sc ii}     &    0   &     1   &     0   \\ 
C\,{\sc i}      &   18   &    15   &    19   \\ 
C\,{\sc ii}     &   30   &    17   &    32   \\ 
C\,{\sc iii}    &   54   &    13   &    32   \\ 
C\,{\sc iv}     &    0   &     1   &     0   \\ 
Mg\,{\sc i}     &   21   &    17   &    31   \\  
Mg\,{\sc ii}    &   16   &    14   &    34   \\ 
Mg\,{\sc iii}   &    0   &     1   &     0   \\ 
Si\,{\sc i}     &   11   &    19   &    29   \\ 
Si\,{\sc ii}    &    5   &    20   &    48   \\ 
Si\,{\sc iii}   &   17   &    17   &    27   \\ 
Si\,{\sc iv}    &    0   &     1   &     0   \\ 
Ca\,{\sc i}     &   21   &     7   &     3   \\ 
Ca\,{\sc ii}    &   31   &    14   &    21   \\ 
Ca\,{\sc iii}   &    0   &    18   &    25   \\ 
Ca\,{\sc iv}    &    0   &     1   &     0   \\ 
O\,{\sc ii}     &   31   &    16   &    26   \\ 
O\,{\sc iii}    &   51   &    21   &    38   \\ 
O\,{\sc iv}     &    0   &     1   &     0   \\          
\hline                               
\end{tabular}
\end{table}

Detailed information about the involved atomic data is provided in the form of a model atom \citep[cf.][]{Rauch:2003p1718}. The model atoms we used for our NLTE calculations are summarised in Table~\ref{tab1}. They were taken from \emph{TMAD}, the T\"ubingen Model Atom Database\footnote{http://astro.uni-tuebingen.de/\raisebox{.2em}{\tiny $\sim$}TMAD/TMAD.html}.

The principal problem for any modelling attempt is posed by the question of what heats the Ca\,{\sc ii} emission line region. It cannot be gravitational energy released through viscosity because the required mass-accretion rate would have to be of the order of $10^{-8}\,$M$_\odot$/yr, which is by many orders of magnitude larger than the accretion rate invoked for the presence of settling metals in DAZ photospheres \citep[$\approx 10^{-15}\,$M$_\odot$/yr,][]{Koester:2006p1769}. A speculation by \cite{Jura:2008p1777} was additional heating by energy dissipation through disc asymmetries, which are driven by some external unseen planet. Alternatively, \cite{2010ApJ...722.1078M} suggested a ``Z\,{\sc ii}'' model in analogy to H\,{\sc ii} regions. In the case of the discs, the metal-dominated material is photoionised and heated (hence the name Z\,{\sc ii}) by absorbing photons from the WD and cools through optically thick emission lines. Given this lack of knowledge, we need to use $T_{\rm{eff}}$, which is a measure of the vertically integrated dissipated energy (Eq.\,\ref{eqmech}), hence the amount of energy
radiated away from the disc surface per unit time and area, as a free parameter.

%__________________________________________________________________
\section{SDSS\,J122859.93+104032.9}\label{sdss}

Our models are tailored to SDSS\,J122859.93+104032.9 (henceforth SDSS\,J1228+1040). This metal-polluted WD was the first one discovered to be surrounded by a gaseous metal disc \citep{Gansicke:2006p444}. It is a DAZ white dwarf with atmospheric parameters $T_{\rm{eff}}$\,=\,$22\,020 \pm 200$\,K and log\,$g$\,=\,$8.24 \pm 0.04$, and the derived stellar mass and radius are $M_{\rm{WD}}$\,=\,$0.77 \pm 0.02$\,M$_\odot$ and $R_{\rm{WD}}$\,=\,$0.0111 \pm 0.0003$\,R$_\odot$.

The Ca\,{\sc ii} infrared triplet ($\lambda\lambda$ 8498, 8542, 8662\,\AA) with double-peak emission line profiles is the hallmark of the gaseous metal discs (Fig.\,\ref{Fig1}). In the case of SDSS\,J1228+1040, \cite{Gansicke:2006p444} measured a peak-to-peak separation of 630\,km/s, i.e. the Keplerian rotation velocity is $v\,\sin i$\,=\,315\,km/s. From a spectral analysis with a kinematical LTE emission model, it was concluded that we see a geometrically thin, optically thick gaseous disc at high inclination ($i=70^\circ$). Two other weak emission features of Fe\,{\sc ii} $\lambda\lambda$ 5018, 5169\,\AA\ were seen by \cite{Gansicke:2006p444}. Subsequent observations by \cite{2010ApJ...722.1078M} failed to detect the Fe\,{\sc ii} $\lambda$ 5018\,\AA\ line, possibly because of the lower signal-to-noise ratio of their high-resolution spectra as compared to the low-resolution spectrum of \cite{Gansicke:2006p444}. On the other hand, very weak emissions from the Ca\,{\sc ii} H and K lines were discovered by \cite{2010ApJ...722.1078M}.

There is a clear asymmetry in the emission strengths of the double-peak line profiles of SDSS\,J1228+1040 \citep{Gansicke:2006p444}. A similar phenomenon is well known from Be star discs \citep{car10} and is ascribed to one-armed spiral waves. In addition, \cite{2010ApJ...722.1078M} observe that the asymmetry in SDSS\,J1228+1040 had changed such that the stronger of the two emission peaks has switched from the red side of the double-peaked emission complex, as seen in \cite{Gansicke:2006p444}, to the blue side. We describe these characteristics in more detail in Sect.\,\ref{sect_var}, where we investigate the temporal variability predicted by our models.

%__________________________________________________________________
\section{Results}\label{results}

\subsection{Parameter study for pure calcium discs}

\begin{figure}
\centering
\includegraphics[width=0.45\textwidth]{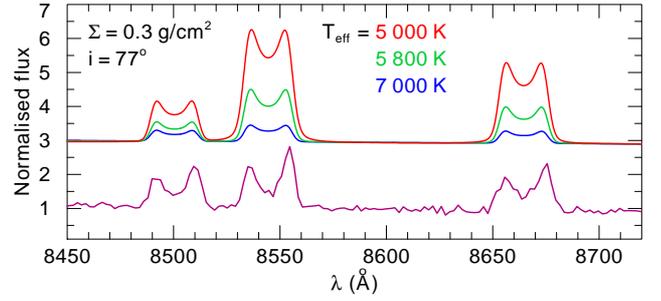}
\caption{Normalised spectra of three pure Ca disc models (top) with different $T_{\rm{eff}}$ compared to the observed spectrum of SDSS1228+1040 taken from SDSS (bottom). The model spectra are shifted vertically for clarity.}
\label{Fig1}
\end{figure}

In a first exploratory study \citep{Werner:2009p373}, we calculated disc models composed only of calcium, with two values for $\mathit{\Sigma}$ and three values for $T_{\rm{eff}}$. The inner and outer disc radii were set to 1.0\,R$_\odot$ and 1.2\,R$_\odot$, respectively.  We found that the emission strength of the Ca\,{\sc ii} triplet decreases with $T_{\rm{eff}}$  increasing from 5\,000\,K to 7\,000\,K (Fig.\,\ref{Fig1}). The reason is the shifting Ca\,{\sc ii}/Ca\,{\sc iii} ionisation balance. A closer comparison of the three line components shows that their emission strengths become equal with increasing $T_{\rm{eff}}$, a behaviour that constrains $T_{\rm{eff}}$. A similar trend is seen when $\mathit{\Sigma}$ is reduced from 0.6\,g/cm$^2$ to 0.3\,g/cm$^2$ at $T_{\rm{eff}}$\,=\,6\,500\,K. We stress that the models have a considerable continuum flux compared to the line-emission peak heights. The relative strength of the profile depression between the double-peaks increases with increasing inclination, the double peak structures become broader. The disc models are optically thin in terms of the Rosseland optical depth.

\begin{figure}
\centering
\includegraphics[width=0.95\columnwidth]{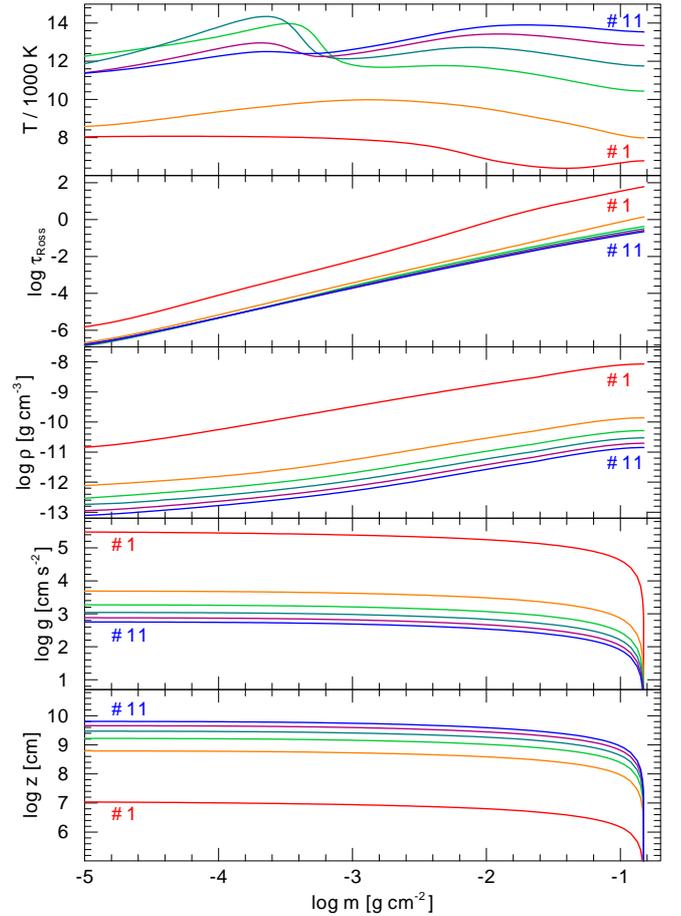}
\caption{Vertical structure of every other of the 11 disc rings along the column-mass density (increasing from outer layers towards the disc's midplane).}
\label{vertstruct}
\end{figure}

The observed spectrum does not show H$_\alpha$ emission. This can be used to determine an upper limit for the hydrogen abundance. We varied the H content (H\,=\,1\%, 0.1\%, 0.01\%, by mass) and found that with an abundance of 1\%, the H$_\alpha$ peak height is comparable to that of the Ca\,{\sc ii} triplet, so would be detectable in the spectrum of SDSS\,J1228+1040.

The effective temperature of the disc is well constrained by three models, with $T_{\rm eff}$\,$\approx$\,5\,800\,K (Fig.\,\ref{Fig1}), but the asymmetry of the line profiles is of course not matched by our symmetric models. The cooler model ($T_{\rm eff}$\,=\,5\,000\,K) is perhaps more favourable because of the larger line-to-continuum emission ratio, while the hotter model ($T_{\rm eff}$\,=\,7\,000\,K) has the advantage that the relative strengths of the three line components are reproduced better.

\begin{figure*}
\centering
\includegraphics[width=0.66\textwidth,angle=-90]{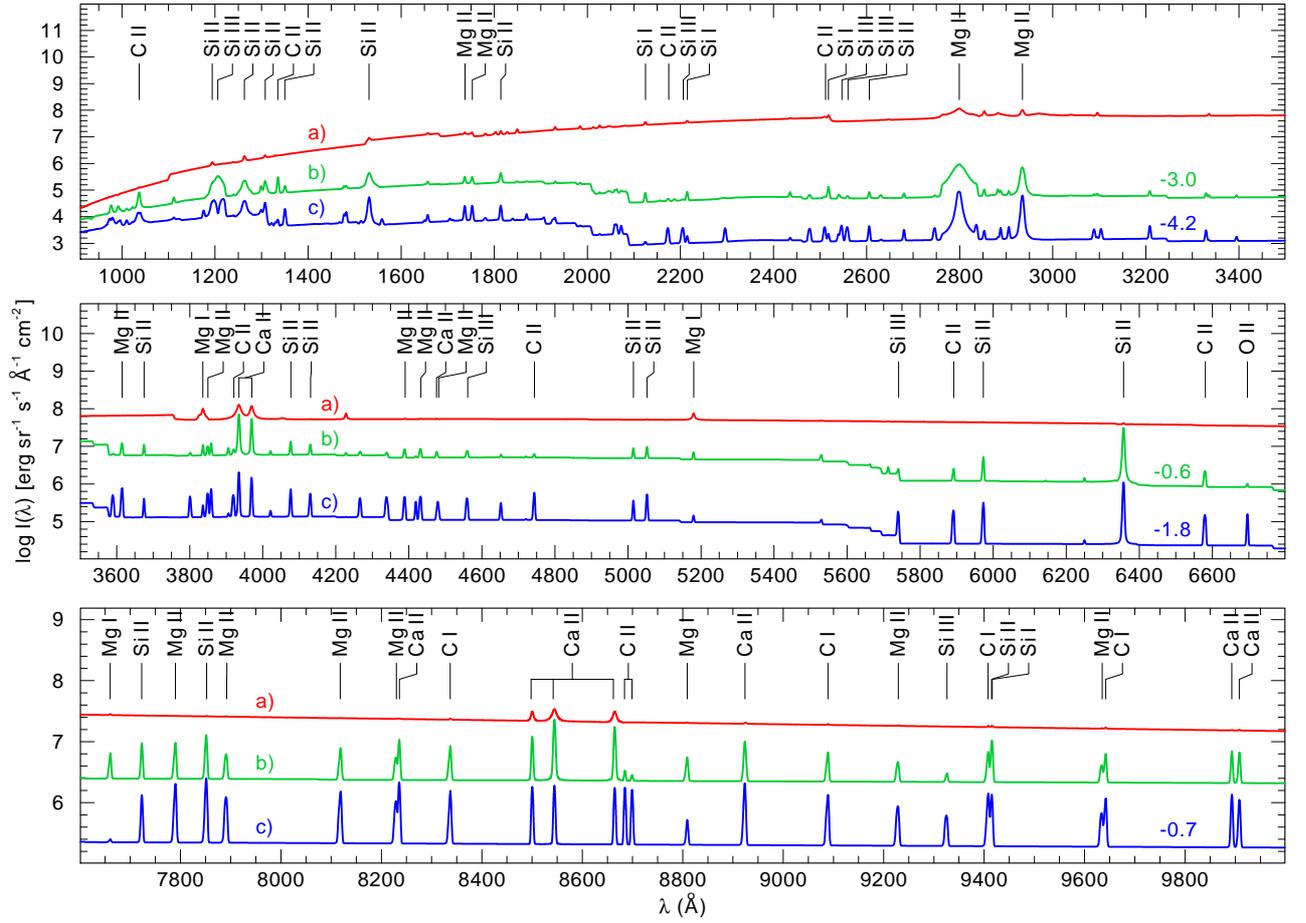}
\caption{Spectra of three (non-rotating) circumstellar rings at $i=77^\circ$ and: (a) 2.0, (b) 71.4, and (c) 136.4\,R$_{\rm WD}$. For clarity, the spectra are vertically shifted by amounts as colour-coded numerically on the plots. With the exception of the infrared Ca\,{\sc ii} triplet and the adjacent C\,{\sc ii} doublet, as well as Ca\,{\sc ii} H~\&~K, no fine structure splitting was taken into account.}
\label{ringspectra}
\end{figure*}

\subsection{O-Si-Mg-C-Ca-H disc models}

In the next step, we expanded the set of considered chemical elements in order to achieve a composition comparable to CI chondrites in the Solar System or a bulk-Earth mixture \citep{Klein:2010p719}. New species included are O, Si, Mg, and C. Iron poses special numerical problems and will be introduced in future work. The initially chosen element abundances are representative of the class of CI chondrites. In detail, they are: H = 10$^{-8}$, C = 4.6, O = 65.5, Mg = 13.5, Si = 15.1, Ca = 1.3 (\% mass fraction).

We investigated the influence of the radial disc extent on the spectrum. At its largest, the disc model consists of 11 rings extending from $R_{\rm{i}}$\,=\,2\,R$_{\rm{WD}}$\,=\,0.022\,R$_\odot$ to an outer radius of $R_{\rm{o}}$\,=\,136\,R$_{\rm{WD}}$\,=\,1.5\,R$_\odot$. The ring radii and effective temperatures are listed in Table~\ref{tab:r}. The model for ring 8 did not converge; its spectrum was set equal to that of ring 9. The entire disc is assumed to have a radially constant surface density of 0.3\,g/cm$^2$.

In Fig.\,\ref{vertstruct} we present the vertical run of temperature, Rosseland optical depth $\tau_{\rm ross}$, mass density, gravity, and geometrical height of the disc rings. Only the inner rings are optically thick. Figure~\ref{ringspectra} shows spectra of three different rings, at 2.0, 71.4,  and 136.4\,R$_{\rm WD}$. They all show emission lines, getting stronger in the outer parts of the disc.

\begin{table}
\caption{Radial position and effective temperature of the 11 concentric rings forming the basic disc.}
\label{tab:r}
\centering
\begin{tabular}{r r r r}
\hline\hline
\noalign{\smallskip}
Ring No.& $R\,/\,10^{10}$\,cm& $R\,/$\,R$_{\rm WD}$& $T_{\rm eff}\,/$\,K \\
\hline
       1&              0.159&                 2.1&             6\,707\\
       2&                1.2&                15.6&             6\,294\\
       3&               2.35&                30.5&             6\,051\\
       4&                3.5&                45.5&             5\,907\\
       5&                4.5&                58.4&             5\,903\\
       6&                5.5&                71.4&             5\,824\\
       7&                6.5&                84.4&             5\,780\\
       8&                7.5&                97.4&                  -\\
       9&                8.5&               110.4&             5\,714\\
      10&                9.5&               123.4&             5\,693\\
      11&               10.5&               136.4&             5\,678\\
\hline
\end{tabular}%
\end{table}%

\begin{figure}
\centering
\includegraphics[width=0.45\textwidth]{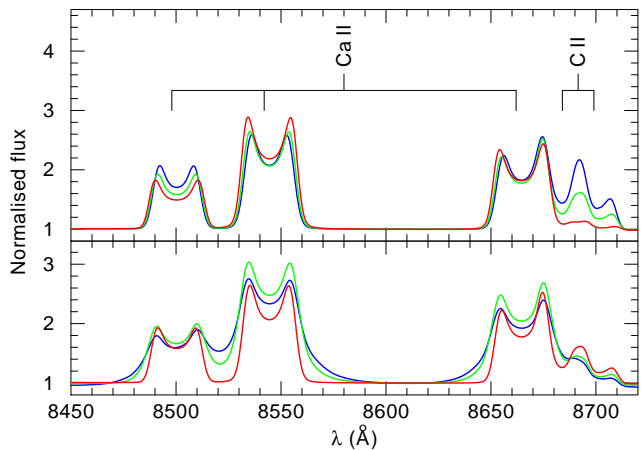}
\caption{Effect of variation in the disc's radial extent at $i=$\,77$^\circ$. Upper panel: the inner radius is fixed at 58\,R$_{\rm{WD}}$ and the outer radius varies: 84, 110, and 136\,R$_{\rm{WD}}$. The C\,{\sc ii} line disappears with decreasing outer radius. Lower panel: The outer radius is fixed at 110\,R$_{\rm{WD}}$ and the inner radius varies: 2, 30, and 58\,R$_{\rm{WD}}$. Broad line wings develop when the inner radius becomes smaller.}
\label{rand}
\end{figure}
\begin{figure}
\centering
\includegraphics[width=0.45\textwidth]{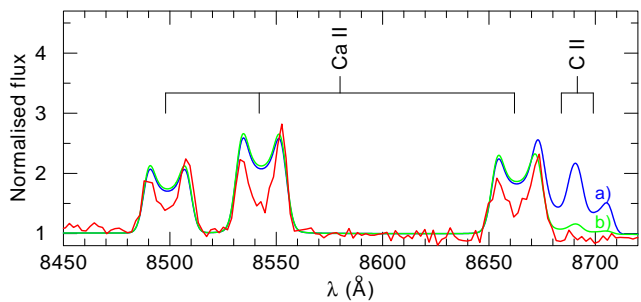}
\caption{Variation in the carbon abundance. Shown are two models (a: 4.6\% and b: 0.46\% by mass) and the observed spectrum (red). The radial disc extent is 58--136\,R$_{\rm{WD}}$, $i=$\,77$^\circ$.}
\label{C2abund}
\end{figure}

We combined the 11 rings to several axisymmetric discs with varying radial extent. Their spectra are shown in Fig.\,\ref{rand}. The dip between the triplet line components of the Ca\,{\sc ii} lines seen in observed spectra can best be reproduced if the disc does not reach too far inwards, resulting in a minimum inner radius of about 60\,R$_{\rm{WD}}$. Variation in the outer disc radius has almost no effect on the Ca\,{\sc ii} triplet, but reducing the outer radius decreases the line strength of a C\,{\sc ii} doublet ($\lambda\lambda$ 8685, 8699\,\AA) near the red component of the Ca\,{\sc ii} triplet. While the lack of observed C\,{\sc ii} emission may suggest a constraint on the outer disc radius, we note that decreasing the carbon abundance by an order of magnitude also removes this feature from the model spectra (see Fig.\,\ref{C2abund}).  Given the carbon-poor nature of the observed discs \citep{rea05,jura09} and polluted WD atmospheres \citep{jura06}, abundance may play a substantial role.

Our model strongly overpredicts the strength of the Ca\,{\sc ii} H~\&~K emission lines. This problem has already been noted in the pure-Ca disc described above \citep{Werner:2009p373}. This problem did not disappear in the present multi-element disc model. In contrast, our model features emission lines from other metals, markedly from Mg\,{\sc ii} that are not observed. Another weakness of our model is the relatively strong continuum flux that would detectable by distorting the WD spectrum. This problem is alleviated when the effective temperature is reduced.

\subsection{Asymmetric disc models}\label{sect_var}

To investigate the asymmetry of the line profiles in the spectra of SDSS\,J1228+1040, we modified our method combining the disc rings to receive a complete disc spectrum. In the surface integrating step, we used only parts of the rings in order to construct a spiral-arm like or an eccentric shape of the disc. The ring segments are still assumed to undergo Keplerian rotation for the calculation of the spectral Doppler shift. Depending on the orientation towards the observer, these non-axisymmetric accretion discs result in asymmetric line profiles that can be compared to the observations.

We performed hydrodynamic calculations with the FARGO code \citep{Masset2000} in order to motivate our particular choice of geometries. This code was originally developed to compute the hydrodynamic evolution of a protoplanetary disc on a fixed polar coordinate system, but it is also suitable for other kinds of sheared fluid discs. A possible scenario could be that an asteroid coming within the tidal radius of the WD is disrupted, forming a locally concentrated debris cloud. Matter then spirals inwards. We start the simulations by putting a gas blob at the tidal radius ($R=137$\,R$_{\rm{WD}}$) onto a circular orbit ($P_{\rm orb}=3.6 \cdot 10^{4}$\,s) around the WD. The gas mass is 7$\cdot 10^{21}$\,g. The initial surface density distribution is Gaussian with a blob radius of 2.8$\cdot 10^{5}$\,km. We chose an open boundary condition and fixed the simulation rim at a value of $R=194$\,R$_{\rm{WD}}$.

\begin{figure}
\centering
\includegraphics[width=0.5\textwidth]{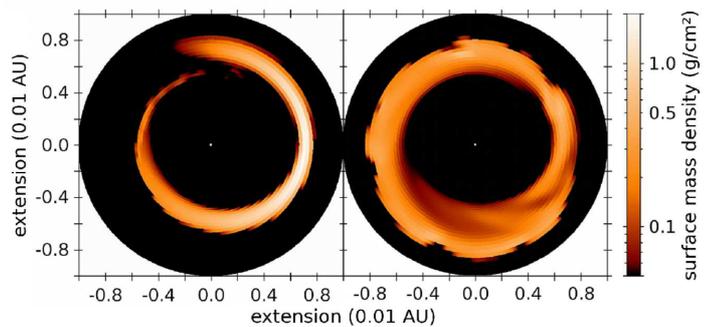}
\caption{Two exemplary structures from the hydrodynamical simulations. The initial gas blob first evolves into a spiral-arm structure (left) and after some time into an eccentric disc (right). The surface density is colour-coded as indicated.}
\label{fargo}
\end{figure}

\begin{figure}
\centering
\includegraphics[width=0.5\textwidth]{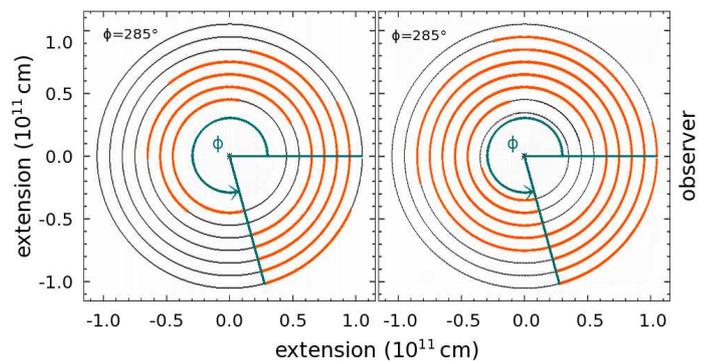}
\caption{Sketch of the two non-axisymmetric geometries used for our spectral models. Left: a spiral-arm like structure. Right: an eccentric accretion disc. Both are composed of disc ring segments (red). For the unmarked ring segments, the emergent flux is set to zero. The observer's position is towards the right.}
\label{geo}
\end{figure}

The Keplerian rotating material gets smeared out into a spiral-arm like structure within a short time of $t$\,=\,1.21$\cdot 10^{4}$\,s. After another $\Delta t$\,=\,3.1$\cdot 10^{4}$\,s,  the spiral is catching up with its own starting point and an eccentric closed disc forms and retains this shape for the rest of the simulation. For both situations the surface mass density is rather homogeneous with $\mathit{\Sigma}\approx$\,0.3\,g/cm$^2$. Examples for both geometries found by the simulations are shown in Fig.\,\ref{fargo}. In Fig.\,\ref{geo} we display the corresponding assembly of our ring segment models.

\begin{figure}
\centering
\includegraphics[width=0.45\textwidth]{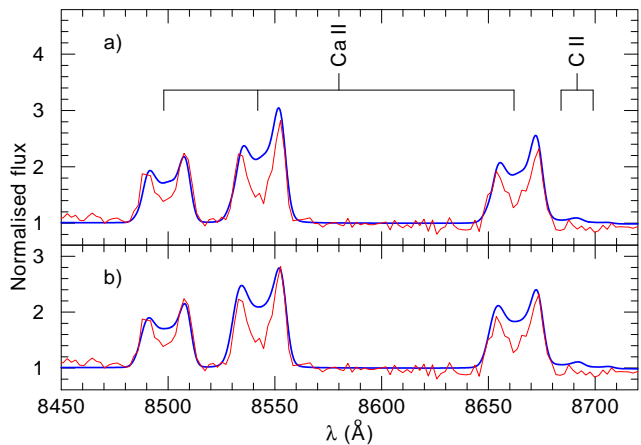}
\caption{Emergent spectrum for the (a) spiral-arm like and (b) eccentric disc structures as shown in Fig.\,\ref{geo}. The modelled profiles (thick lines) display an asymmetry as observed (thin line). For the spiral arm the radial extent is 58--136\,R$_{\rm{WD}}$, whereas in the case of the eccentric disc it is 45--136\,R$_{\rm{WD}}$. For both, $i=$\,77$^\circ$, $\phi$\,=\,285$^\circ$, and $\mathit{\Sigma}$\,=\,0.3\,g/cm$^2$.}
\label{deformspek}
\end{figure}

The upper panel of Fig. \ref{deformspek} shows the resulting spectrum in the case of a spiral-arm like shape with a minimum radius of 58\,R$_{\rm{WD}}$ and a maximum radius of 136\,R$_{\rm{WD}}$ for an observer's position according to Fig.\,\ref{geo} (left). The strength of the Ca\,{\sc ii} lines and the asymmetry of the blue and red parts of each line component are well reproduced. Only the dip between the Doppler-shifted parts of each line are not as deep as in the observation. The mass of such a spiral arm would be about $3.9\cdot 10^{21}$\,g. In the lower panel of Fig.\,\ref{deformspek}, the resulting spectrum of an eccentric disc for an observer's position according to Fig.\,\ref{geo} (right) is shown. The fit quality to the observed spectrum is similar to the spiral-arm case. Such an eccentric disc would have a mass of about $6.3\cdot 10^{21}$\,g.

\begin{figure}
\centering
\includegraphics[width=0.5\textwidth]{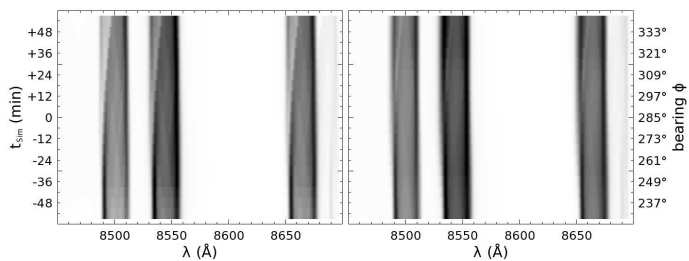}
\caption{Time series of the computed Ca\,{\sc ii} triplet emission, covering 100\,min. For the spiral-arm geometry (left panel), a change in the line-profile asymmetry evolves very fast, whereas for the eccentric disc (right) it takes much longer.}
\label{timeserie}
\end{figure}

\subsection{Time variability}

\citet{Gansicke:2006p444} present two time series of spectra from SDSS\,J1228+1040 taken on June 30 and July 1, 2006. They indicate that the asymmetry of the Ca\,{\sc ii} line profiles did not change during the two 100\,min observations. To investigate this with our models, we assumed a Keplerian rotating spiral arm and an eccentric disc with $\omega$\,=\,0.01$^\circ$/s, and calculated synthetic spectra in steps of $\Delta\phi$\,=\,6$^\circ$ concerning the orientation towards the observer, starting at $\phi$\,=\,231$^\circ$ and centred on 285$^\circ$, for which we found the best fit to the emission lines. The resulting spectral time series are shown in Fig.\,\ref{timeserie}, covering 100\,min in order to be comparable with the lower right part of Fig.\,1 in \citet{Gansicke:2006p444}. In the case of a spiral-arm like geometry, the asymmetry of the line profile would change significantly within this time interval (Fig.\,\ref{timeserie}, left panel). The separation of the maxima of the double peaks decreases, and at the same time the Ca\,{\sc ii} line asymmetry reverses in wavelength. In contrast, for the eccentric disc the relative strength of the asymmetry changes quite slowly (Fig.\,\ref{timeserie}, right) and the double-peak separation remains almost constant, which is in better agreement with the non-variable observations. On the other hand, \cite{2010ApJ...722.1078M} found a switch of the asymmetry on a longer timescale, between the \citeauthor{Gansicke:2006p444} observations in 2006 and their own in 2008.

%__________________________________________________________________
\section{Summary and conclusion}

We performed non-LTE modelling of gaseous metal discs around WDs in order to study their spectral signatures in comparison to observations of SDSS\,J1228+1040. The modelling was done in three steps.

At first, pure calcium models were constructed to constrain the disc characteristics by fitting the observed infrared Ca\,{\sc ii} emission triplet. Qualitatively good fits can be obtained with a geometrically and optically thin, Keplerian viscous gas disc ring at a distance of 1.2\,R$_\odot$ from the WD, with $T_{\rm{eff}}$\,$\approx$\,5\,800\,K and a low surface mass density $\mathit{\Sigma}$\,$\approx$\,0.3\,g/cm$^2$. The disc is hydrogen-deficient (H\,${<}$\,1\% by mass), and it is located within the tidal disruption radius ($R_{\rm{tidal}}$\,=\,1.5\,R$_\odot$).

In the second step, we constructed axisymmetric disc models composed of elements in Chondritic abundance, namely C, O, Si, Mg, and Ca. We found that the inner radius of the observed, emitting, Ca\,{\sc ii} gaseous component of the disc is well constrained at $\ge$\,0.65\,R$_\odot$\,=\,58\,R$_{\rm{wd}}$. The outer radius can be constrained by the emission strength of a C\,{\sc ii} doublet ($\lambda\lambda$ 8685, 8699\,\AA) that is not seen in the observations. An alternative explanation could be a reduced carbon abundance, which would be a hint that the disc in SDSS\,J1228+1040 has a bulk-Earth like composition instead of a CI chondritic one.

In the third step, we investigated asymmetric disc structures by assuming spiral-arm and eccentric disc shapes as suggested by hydrodynamical simulations. Both geometries can qualitatively explain the asymmetry observed in the double-peak line profiles well. An investigation of the time variability of the computed line profiles suggests that the eccentric disc model displays less significant variability than the spiral-arm geometry. Considering the current observational material, the eccentric disc model is more realistic. Mass estimates for the circumstellar gas material using the two geometric models results in 3--6\,$\cdot 10^{21}$\,g, which are equivalent to the mass of a 135-km diameter Solar System asteroid.

%__________________________________________________________________
\begin{acknowledgements}
We thank Tobias M\"uller from the Computational Physics group at our institute for providing us with the FARGO simulations. S.H. and T.R. are supported by DFG (grant We1312/37-1) and DLR (grant 50\,OR\,0806), respectively.
\end{acknowledgements}

%__________________________________________________________________
\begin{figure}
\centering
\includegraphics[width=0.30\textwidth]{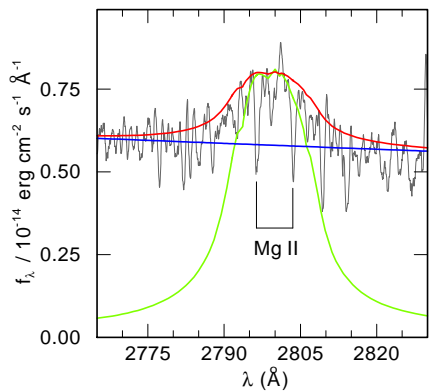}
\caption{Mg\,{\sc ii} resonance doublet in the HST/COS spectrum, overplotted with our (rotating) disk model and a pure-H WD model, plus the co-added model. Fine-structure splitting of the line is considered in the disk model.}
\label{mg2}
\end{figure}
\noindent
\emph{Note added in proof.} Following the acceptance of our paper, an HST/COS observation of SDSS\,J1228+1040 became public (dataset LB5Z040, observation date 2010-02-19). This is the first available UV observation. The spectrum shows a strong emission line of Mg\,{\sc ii} that our disk model, as displayed in Fig.\,\ref{ringspectra}, predicts to be the strongest UV line. This discovery is essential because Mg is the third element (after Ca and Fe) that is observed in the metal disk, supporting the idea of ground planetary material. Figure~\ref{mg2} shows the observed Mg\,{\sc ii} resonance doublet ($\lambda\lambda$ 2796.35, 2803.53\,\AA) in emission with central absorption components of photospheric origin. Overplotted is a pure-H WD model scaled to fit the continuum and our disk model (including all rings 5--11 as listed in Table~\ref{tab:r}, $i=77^\circ$) scaled arbitrarily, plus the co-added WD+disk model spectrum. The optical depth in the Mg\,{\sc ii} line varies across the disk and has a maximum of $\tau \sim 10^6$ in the outermost disk region.

\bibliographystyle{aa}

\end{document}